\documentclass[aps,prb,twocolumn,reprint]{revtex4-1}
\usepackage[T1]{fontenc} 
\usepackage{amsmath}
\usepackage{amsfonts}
\usepackage{color}
\usepackage{xcolor}
\usepackage{ucs}
\usepackage{graphicx,bm,hyperref}
\usepackage[caption=false]{subfig}
\usepackage{breqn}

\makeatletter
\let\cat@comma@active\@empty
\makeatother

\newcommand{\angstrom}{\textup{\AA}}
\newcommand{\ep}{\epsilon}

\title{Thermofield Theory for Finite Temperature Coupled Cluster}

\begin{document}
\title{Thermofield theory for finite-temperature coupled cluster}
\author{Gaurav Harsha}
\affiliation{Department of Physics and Astronomy, Rice University, Houston TX 77005}
\author{Thomas M. Henderson}
\affiliation{Department of Physics and Astronomy, Rice University, Houston TX 77005}
\affiliation{Department of Chemistry, Rice University, Houston TX 77005}
\author{Gustavo E. Scuseria}
\affiliation{Department of Physics and Astronomy, Rice University, Houston TX 77005}
\affiliation{Department of Chemistry, Rice University, Houston TX 77005}

\begin{abstract}
    We present a coupled cluster and linear response theory to compute properties of many-electron systems at non-zero temperatures. For this purpose, we make use of the thermofield dynamics, which allows for a compact wavefunction representation of the thermal density matrix, and extend our recently developed framework [J. Chem. Phys. 150, 154109 (2019)] to parameterize the so-called thermal state using an exponential ansatz with cluster operators that create thermal quasiparticle excitations on a mean-field reference. As benchmark examples, we apply this method to both model (one-dimensional Hubbard and Pairing) as well as \emph{ab-initio} (atomic Beryllium and molecular Hydrogen) systems, while comparing with exact results.
\end{abstract}

\maketitle

\section{\label{sec1}Introduction}
Temperature is an important parameter of physical systems. For many problems, the temperature scale of interest is far below the optical gap (or the excitation energies). For example, typical electronic excitation energies in molecules are of the order of a few electron volts (or $\mathrm{eV}$) , which is much larger than room temperature ($\sim 25 \mathrm{meV}$). In such problems where we are interested only in electronic degrees of freedom, it suffices to know the ground electronic state and perhaps a few low-lying excited states. To access these states, we solve the time-independent Schr\"{o}dinger equation. As is well known,\cite{dirac_quantum_1929} this is a very complicated problem and one generally relies on a series of approximate methods such as Hartree-Fock, perturbation theory, configuration interaction (CI), coupled cluster theory\cite{crawford_introduction_2000,bartlett_coupled-cluster_2007} (CC), Monte Carlo methods,\cite{hammond_monte_1994, foulkes_quantum_2001, zhang_quantum_2003, zhang_quantum_2004, al-saidi_auxiliary-field_2006} among many more.\cite{white_density-matrix_1993,ostlund_thermodynamic_1995,georges_dynamical_1996}

There are, however, many interesting problems and applications where one may be interested in temperature scales that are comparable or even larger than the excitation gap. Examples include metallic compounds with small gap that host an unconventional superconductivity persisting at relatively high temperatures,\cite{lee_doping_2006} ultra-cold chemistry,\cite{balakrishnan_perspective_2016,bohn_cold_2017} geochemical processes which generally involve very high temperatures and pressure,\cite{guillot_interiors_1999} etc. In these problems, we can no longer make do with a few electronic states and must evaluate properties as thermal averages weighted over an appropriate ensemble of states. For a system in thermal equilibrium at inverse temperature $\beta$, the ensemble of choice is generally canonical or grand-canonical and the expectation value of an observable $A$ is defined by
\begin{equation}
    \langle A \rangle_\beta = \frac{1}{\mathcal{Z}} \mathrm{Tr} ( A \rho ),
    \label{trace-avg}
\end{equation}
where $\rho$ is the thermal density matrix, and $\mathcal{Z}$ is the partition function, given by
\begin{equation}
    \rho = e^{-\beta H'},  \quad \mathcal{Z} = \mathrm{Tr} ( \rho ),
\end{equation}
with $H' = H$ for the canonical ensemble and $H' = H-\mu N$ for the grand canonical ensemble, where $\mu$ is the chemical potential. In this paper, we shall work explicitly with the grand canonical ensemble.

Exactly computing $\rho$ (or equivalently $\mathcal{Z}$) requires information about the entire spectrum of the Hamiltonian, which is far from feasible. Accordingly, just as for zero temperature, a series of approximate methods is needed. Several methods have been proposed over the years to evaluate thermal averages of operators.\cite{mermin_stability_1963,sokoloff_consequences_1967,lichtenstein_finite-temperature_2001,verstraete_matrix_2004,feiguin_finite-temperature_2005, white_minimally_2009, stoudenmire_minimally_2010,pittalis_exact_2011, pizorn_real_2014, hermes_finite-temperature_2015, czarnik_variational_2016, santra_finite-temperature_2017, zgid_finite_2017, claes_finite-temperature_2017, white_time-dependent_2018, hummel_finite_2018} Most of these methods can be broadly categorized into deterministic methods such as diagrammatic perturbation theory based on the Matsubara formalism,\cite{matsubara_new_1955} and stochastic quantum Monte Carlo methods.
Wavefunction methods are particularly convenient in the study of zero-temperature ground-state properties of finite sized systems and clearly, their thermal equivalents are highly desirable. However, the development of such thermal wavefunctions methods has been rather challenging, primarily because the thermal density matrix cannot be expressed in terms of a single wavefunction in the original Hilbert space. Nevertheless, several such methods have been introduced over the years such as the Ancilla density matrix renormalization group\cite{verstraete_matrix_2004,feiguin_finite-temperature_2005} and finite-temperature perturbation theories,\cite{he_finite-temperature_2014,santra_finite-temperature_2017} among many more.

Given its features, especially size-extensivity and success with weakly correlated systems, the coupled cluster ansatz is an ideal candidate to study finite-temperature properties.
A thermal analogue of the CC method\cite{sanyal_thermal_1992,sanyal_systematic_1993,mandal_thermal_1998,mandal_finite-temperature_2003} was proposed by Mukherjee \emph{et. al.} and has been further elaborated recently in independent works by White \emph{et al.}\cite{white_time-dependent_2018} and Hummel.\cite{hummel_finite_2018} This formulation uses a thermal Wick's theorem to compactly represent the imaginary time evolution operator as a \emph{thermal normal ordered} exponential of some cluster operator and a number.

In this paper, we present an alternative approach to thermal coupled cluster based on the thermofield dynamics (TFD),\cite{matsumoto_thermo_1983,semenoff_functional_1983,umezawa_methods_1984,evans_heisenberg_1992} using a framework we recently explored in Ref.~\citenum{harsha_thermofield_2019}. Thermofield dynamics provides a convenient way to represent the thermal density matrix via a wavefunction which evolves in temperature according to the imaginary-time evolution Schr\"odinger equation. Undoubtedly, TFD has the potential to study many-electron systems in quantum chemistry and condensed-matter physics.\cite{suzuki_thermo_1985, hatsuda_mean_1989, walet_thermal_1990, de_vega_thermofield-based_2015,borrelli_quantum_2016, nocera_symmetry-conserving_2016, chen_finite_2017, borrelli_simulation_2017, wu_variational_2018} A connection between thermofield formalism and thermal cluster-cumulant theory was also explored by Mukherjee \emph{et. al.} in Ref.~\citenum{sanyal_systematic_1993}. Here, we parametrize this so-called thermal wavefunction as an exponential ansatz to integrate the evolution equation.

\section{\label{sec2}Coupled Cluster Theory}

The coupled cluster method is one of the most widely used methods in quantum chemistry. Introduced first in nuclear physics by Coester and K\"{u}mmel\cite{coester_bound_1958, coester_short-range_1960} and later reformulated for electronic structure theory by \v{C}i\v{z}ek and Paldus,\cite{cizek_correlation_1971} it uses an exponential wavefunction ansatz
\begin{equation}
    |\Psi \rangle = e^{T} |\Phi \rangle,
\end{equation}
to solve the time-independent Schr\"odinger equation. Here $|\Phi\rangle$ is some Hartree-Fock (HF) Slater determinant reference and $T$ contains particle-hole excitations defined on this reference state,
\begin{subequations}
    \begin{align}
        T &= T_1 + T_2 + \ldots, \\
        T_1 &= \sum_{i,a} t_i^a c^\dagger_a c_i,\\
        T_2 &= \frac{1}{4} \sum_{i,j,a,b} t_{ij}^{ab} c^\dagger_a c^\dagger_b c_j c_i, \ldots
    \end{align}
\end{subequations}
where we have followed the standard notation for labeling orbital indices, i.e., occupied orbitals are denoted by indices $i,j,k,\ldots $, while unoccupied orbitals are denoted by $a, b, c, \ldots$. Since the wavefunction $\vert \Psi \rangle$ is constructed as excitations on a single Slater determinant $\vert \Phi \rangle$, the traditional CC is also known as single-reference CC. In what follows, unless explicitly mentioned, we consider only the single-reference CC. The equations for the ground state energy and the unknown $t$-amplitudes can be obtained by left-projecting the Schr\"odinger equation
\begin{equation}
    \bar{H} \left| \Phi \right \rangle = E \left | \Phi \right \rangle, \quad \bar{H} = e^{-T} H e^{T}
    \label{schro-eq}
\end{equation}
with various Slater determinants. For example, if the cluster operator $T$ is truncated to single and double excitations only (CCSD), the energy and amplitude equations are obtained by solving the following equations
\begin{align}
    E &= \left \langle \Phi \right \rvert \bar{H} \left \lvert \Phi \right \rangle, \\
    0 &= \left \langle \Phi_{i}^{a} \right \rvert \bar{H} \left \lvert \Phi \right \rangle, \\
    0 &= \left \langle \Phi_{ij}^{ab} \right \rvert \bar{H} \left \lvert \Phi \right \rangle,
\end{align}
where $\langle \Phi_{i}^{a} \vert$ and $\langle \Phi_{ij}^{ab} \vert $ are singly- and doubly-excited slater determinants.

As introduced so far, CC describes a correlated ansatz for the \textit{ket} wavefunction. In order to compute expectation values other than that of the Hamiltonian, one also needs a correlated \textit{bra} state. A linear response wavefunction is generally employed for this purpose,\cite{arponen_variational_1983,helgaker_analytical_1988} i.e. one makes the energy functional
\begin{equation}
    E = \langle \Phi \vert (1 + Z) e^{-T} H e^{T} \vert \Phi \rangle
    \label{lin-resp-cc-func}
\end{equation}
stationary with respect to $T$ and $Z$, where
\begin{subequations}
    \label{Z-operator}
    \begin{align}
        Z &= Z_1 + Z_2 + \ldots, \\
        Z_1 &= \sum_{i,a} z_i^{a} c_i^\dagger c_a, \\
        Z_2 &= \frac{1}{4}\sum_{i,a} z_{ij}^{ab} c_i^\dagger c_j^\dagger c_b c_a.
    \end{align}
\end{subequations}
By realizing that the cluster operator $T$ and CI operator $Z$ are composed of particle-hole excitation and de-excitation operators respectively, the \textit{bra} state can be re-written as an explicit CI wavefunction
\begin{equation}
    \langle \Psi'\vert = \langle \Phi  \vert (1 + Z) e^{-T} = \langle \Phi \vert (1 + W) e^{w_0},
    \label{ci2cc}
\end{equation}
where $w_0$ is a constant and $W$ has the same operator-form as $Z$.

A similar formulation known as equation of motion CC\cite{rowe_equations--motion_1968} (EOM-CC) can be used for excited states. Coupled cluster gives highly accurate results for weakly correlated systems and its success can be attributed to the polynomial computational scaling ($\mathcal{O}(N^{6})$ for CCSD) as well as to the fact that the computed properties are size-extensive (scale linearly with the particle number) in the thermodynamic limit.

\section{\label{sec3}Thermofield Dynamics}
Thermofield dynamics is a real-time thermal field theory that treats both time and temperature on equal footing, and was proposed as an alternative to the Keldysh approach in the Matsubara imaginary time formalism. It provides a prescription for purification of the thermal density matrix, allowing us to construct a wavefunction $\vert \Psi(\alpha, \beta) \rangle$, with $\alpha=\beta \mu$, generally known as the \textit{thermal vacuum}, \textit{thermofield double state} or simply as the \textit{thermal state}, so that the trace over an ensemble of states in Eq.~\ref{trace-avg} can be replaced by an expectation value over this wavefunction, i.e.
\begin{equation}
    \langle A \rangle = \frac{ \langle \Psi (\alpha, \beta) \vert A \vert \Psi (\alpha, \beta) \rangle }{ \langle \Psi (\alpha, \beta) \vert \Psi (\alpha, \beta) \rangle }.
    \label{tfd-exp}
\end{equation}
That the thermal state $\vert \Psi (\alpha, \beta) \rangle$ cannot be a pure state in the physical Hilbert space $\mathcal{H}$ is easily established since the density matrix $\rho$ represents mixed states to begin with. In TFD, one therefore introduces a fictitious, conjugate copy of the original Hilbert space, known as the \textit{tilde}-conjugate space or $\tilde{\mathcal{H}}$, and the thermal state is then defined in the doubled space $\mathcal{H} \otimes \tilde{\mathcal{H}}$ as
\begin{equation}
    \vert \Psi (\alpha, \beta) \rangle = e^{(\alpha N-\beta H)/2} \vert \mathbb{I} \rangle, \quad \vert \mathbb{I} \rangle = \sum_m \vert m, \tilde{m} \rangle,
    \label{thermal-vacuum}
\end{equation}
where $H$ is the Hamiltonian, $\{\vert m \rangle \}$ is some orthonormal basis in the Hilbert space $\mathcal{H}$ and $\vert \tilde{m} \rangle $ is the tilde-state corresponding to $\vert m \rangle$, while we have used the shorthand notation $\vert m, \tilde{m}\rangle = \vert m \rangle \otimes \vert \tilde{m} \rangle$. The state $\vert \mathbb{I} \rangle$ is the state with maximal entanglement between $\mathcal{H}$ and $\tilde{\mathcal{H}}$ and is the exact thermal state at infinite temperature (i.e. $\beta=0$) and $\alpha=0$. Moreover, it is invariant under any transformation of the basis. Further details about the TFD formalism and the structure of the \textit{tilde}-conjugate space $\tilde{\mathcal{H}}$ can be found in Ref.~\onlinecite{harsha_thermofield_2019} and the references therein.

By construction, the thermal state satisfies the following imaginary time evolution equations,
\begin{subequations}
    \label{beta-mu-schro}
    \begin{align}
        \frac{ \partial }{\partial \beta} \vert \Psi (\alpha, \beta) \rangle &= -\frac{1}{2} H \vert \Psi (\alpha, \beta) \rangle.
        \label{beta-schro-eq}
        \\
        \frac{ \partial }{\partial \alpha} \vert \Psi (\alpha, \beta) \rangle &= \frac{1}{2} N \vert \Psi (\alpha, \beta) \rangle.
        \label{alpha-schro-eq}
    \end{align}
\end{subequations}
One can solve for $\vert \Psi (\alpha, \beta)\rangle $ by integrating Eq.~\ref{beta-mu-schro}, generally starting from $(\alpha,\beta)=(0,0)$, where the initial thermal state is known exactly.
Exactly evolving the thermal state $\vert \Psi (\alpha, \beta)\rangle$ is equivalent to the exact diagonalization of the density matrix $\rho$. Clearly, approximations need to be introduced in the process.

The simplest approximation that we can invoke is to use the mean-field Hamiltonian $H_0$ instead of $H$. For a many-electron system, the state $\vert \mathbb{I}\rangle$ can be expressed in terms of single-particle Fock states,
\begin{equation}
    \vert \mathbb{I} \rangle = \prod_{p\in \mathrm{levels}} \left ( \vert 0,\tilde{0}\rangle_p + \vert 1,\tilde{1}\rangle_p \right ),
\end{equation}
where $\vert 0 \rangle_p$ and $\vert 1 \rangle_p$ mean that the orbital $p$  is empty or occupied, respectively.
Moreover, if we chose to work with the eigen-basis of $H_0$, i.e.
\[
    H_0 = \sum_p \epsilon_p c^\dagger_p c_p
\]
the normalized mean-field thermal state can be written as
\begin{equation}
    \vert 0 (\alpha, \beta) \rangle = \prod_{p \in \mathrm{levels}} \left ( x_p \vert 0, \tilde{0} \rangle_p + y_p \vert 1, \tilde{1} \rangle \right ),
    \label{mf-thermal-state}
\end{equation}
where $x_p$ and $y_p$ are related to the Fermi-Dirac statistics,
\begin{subequations}
    \begin{align}
        x_p &= \frac{1}{\sqrt{1 + e^{(\alpha - \beta \epsilon_p)}}}, \\
        y_p &= \frac{ e^{(\alpha - \beta \epsilon_p)/2} }{ \sqrt{1 + e^{(\alpha - \beta \epsilon_p)}} },
    \end{align}
\end{subequations}
with $x_p^2 + y_p^2=1$. The  mean-field thermal state in Eq.~\ref{mf-thermal-state} allows us to introduce a thermal Bogoliubov transformation,
\begin{equation}
    \begin{bmatrix}
        a_p  \\
        \tilde{a}_p^\dagger
    \end{bmatrix} =
    \begin{bmatrix}
        x_p & -y_p\\
        y_p & x_p
    \end{bmatrix}
    \begin{bmatrix}
        c_p\\
        \tilde{c}_p^\dagger
    \end{bmatrix},
    \label{hfb-transformation}
\end{equation}
such that
\[
    a_p \vert 0 (\alpha, \beta) \rangle = 0 = \tilde{a}_p \vert 0(\alpha, \beta) \rangle.
\]
Correlated methods can be built with either $\vert \mathbb{I} \rangle$ or $\vert 0(\alpha, \beta) \rangle$ as the reference, while integrating Eq.~\ref{beta-mu-schro}. As we have discussed in Ref~\citenum{harsha_thermofield_2019}, the former choice, which we call the fixed-reference formalism, performs well only in the vicinity of $\beta=0$, while the latter, called the covariant formalism, yields accurate results for the entire range of $\beta$. In the next section, we present details to integrate Eq.~\ref{beta-mu-schro} through the CC ansatz in the covariant formalism. We have also included a short discussion on the fixed-reference approach in Appendix~\ref{app2:fixed}. For brevity of notation, we will henceforth refer the  mean-field thermal state $\vert 0(\alpha, \beta)\rangle$ by $\vert \Psi_0 \rangle$.

\section{\label{sec4}Thermal Coupled Cluster}

As explained in Eq.~\ref{lin-resp-cc-func}, the CC expectation value of any operator $A$ can be evaluated as an asymmetric expectation value,
\begin{equation}
    \langle A \rangle_{\mathrm{CC}} = \frac{ \langle \Psi' \vert A \vert \Psi \rangle }{
        \langle \Psi' \vert \Psi \rangle
    },
\end{equation}
where both the \textit{ket} $\vert \Psi \rangle$ and the \textit{bra} $\langle \Psi' \vert$ states are approximations to the same thermal state, and consequently evolve according to Eq.~\ref{beta-mu-schro} and its adjoint respectively.

Given that the Bogoliubov transformation in Eq.~\ref{hfb-transformation} is BCS-like, i.e. orbitals $p$ (in $\mathcal{H}$) and $\tilde{p}$ (in $\mathcal{\tilde{H}}$) in the mean-field thermal state (Eq.~\ref{mf-thermal-state}) are coupled in just the same way as two opposite momentum single-particle levels in a BCS-wavefunction, we parametrize the \textit{ket} state as an exponential of quasiparticle creation operators\cite{henderson_quasiparticle_2014} acting on an $\alpha$- and $\beta$-dependent mean-field thermal reference, $\vert \Psi_0 \rangle$, defined in Eq.~\ref{mf-thermal-state},
\begin{subequations}
    \begin{align}
        \vert \Psi \rangle &= e^{S(\alpha, \beta)} \vert \Psi_0 \rangle,
        \label{covar-cc-ansatz}
        \\
        S &= s_0
        + \sum_{p,q} s_{pq} a^\dagger_p \tilde{a}^\dagger_p
        + \frac{1}{(2!)^2} \sum_{p,q,r,s} s_{pqrs} a^\dagger_p a^\dagger_q \tilde{a}^\dagger_s \tilde{a}^\dagger_r
        + \ldots
        \label{covar-cc-cluster}
    \end{align}
\end{subequations}
On the other hand, the \textit{bra} state, as in the traditional CC formalism, is approximated as a linear CI-like wavefunction, i.e.
\begin{equation}
    \langle \Psi'\vert = \langle \Psi_0 \vert \: (1 + Z)e^{z_0}e^{-S},
\end{equation}
and as explained in Eq.~\ref{ci2cc}, it can be expressed as an effective CI wavefunction
\begin{subequations}
    \begin{align}
        \langle \Psi'\vert &= \langle \Psi_0 \vert (1 + W) e^{w_0}, \\
        W &= \sum_{p,q} w_{pq} \tilde{a}_q a_p + \frac{1}{4} \sum_{p,q,r,s} w_{pqrs} \tilde{a}_r \tilde{a}_s a_q a_p + \ldots
    \end{align}
\end{subequations}
For both the \textit{bra} and the \textit{ket} states, the reference $\vert \Psi_0 \rangle$ evolves continuously as we evolve the Schr\"odinger Eq.~\ref{beta-mu-schro} (hence the name `covariant'). Accordingly, both the amplitudes ($s_{pq}$, $w_{pq}$, etc.) and the quasiparticle operators $a^\dagger, \tilde{a}^\dagger$ carry $\alpha$- and $\beta$-dependence.
The $\alpha$- and $\beta$-evolution of $\langle \Psi' \vert$, a thermal CI wavefunction, is governed by
\begin{subequations}
    \label{cov-ciAll-beta}
    \begin{align}
        \langle \Psi_0 \vert \left ( \frac{\partial W}{\partial \alpha} + (1 + W) \frac{\partial w_0}{\partial \alpha} \right )
        &= \frac{1}{2} \langle \Psi_0 \vert  N_{\textrm{CI}},
        \\
        \langle \Psi_0 \vert \left ( \frac{\partial W}{\partial \beta} + (1 + W) \frac{\partial w_0}{\partial \beta} \right )
        &= -\frac{1}{2} \langle \Psi_0 \vert H_{\textrm{CI}},
    \end{align}
\end{subequations}
where $H_{\textrm{CI}}$ and $N_{\mathrm{CI}}$ are effective CI Hamiltonian and Number operators respectively, and are given by
\begin{align*}
    H_{\textrm{CI}} &= (1 + W)\,H - H_0\,(1 + W) ,
    \\
    N_{\textrm{CI}} &= W\, N - N\, W.
\end{align*}
A detailed discussion on thermal CI and the derivation of these equations can be found in Ref.~\onlinecite{harsha_thermofield_2019}.

For the evolution of $\vert \Psi \rangle$, substituting the CC ansatz from Eq.~\ref{covar-cc-ansatz} into the Schr\"odinger Eq.~\ref{beta-mu-schro} gives
\begin{subequations}
    \label{covar-cc-gov}
    \begin{align}
        e^{-S} \left ( \frac{\partial}{\partial \alpha} e^{S} \right )\: \vert \Psi_0 \rangle &=
        \frac{1}{2} \left ( e^{-S} N e^{S} - N \right ) \: \vert \Psi_0 \rangle,
        \\
        e^{-S} \left ( \frac{\partial}{\partial \beta} e^{S} \right )\: \vert \Psi_0 \rangle &=
        - \frac{1}{2} \left ( e^{-S} H e^{S} - H_0 \right ) \: \vert \Psi_0 \rangle.
    \end{align}
\end{subequations}
The evolution equations for the amplitudes can be obtained by left projecting Eq.~\ref{covar-cc-gov} with the respective determinants.

\subsection{Wilcox identity}
The process of reducing Eq.~\ref{covar-cc-gov} to evolution equations for the amplitudes is complicated by the fact that the derivative of the cluster operator does not commute with the operator itself, i.e.
\[
    \left [ \frac{\partial S}{\partial x} , S \right ] \neq 0,
\]
where $x = \alpha, \beta$. The derivative of the exponential cluster operator is appropriately performed by making use of the Wilcox identity,\cite{wilcox_exponential_1967} which states that the derivative of the exponential of an operator $M$ with respect to some parameter $\lambda$ can be evaluated as
\begin{equation}
    \frac{\partial }{\partial \lambda}e^{M(\lambda)} = \int_0^1 dy \: e^{(1-y)M} \frac{\partial M}{\partial \lambda} e^{y M}.
    \label{wilcox}
\end{equation}
With this, the left-hand side of Eq.~\ref{covar-cc-gov} becomes
\begin{subequations}
    \begin{align}
        e^{-S} \left ( \partial_x \: e^{S} \right ) &= \int_0^1 dy \: e^{-y S} \: (\partial_x S) \:  e^{y S}, \label{wilcox-demo-1}
        \\
        &= (\partial_x S) + \frac{1}{2!} [ (\partial_x S), S ] \nonumber
        \\
        & \quad + \frac{1}{3!} \left[ [(\partial_x S), S], S \right] + \ldots, \label{wilcox-demo-2}
    \end{align}
\end{subequations}
where we have used the shorthand $\partial_x$ for $\partial/\partial x$, and made use of the Baker-Campbell-Hausdorff expansion in going from Eq.~\ref{wilcox-demo-1} to \ref{wilcox-demo-2}. Finally, breaking the derivative $\partial_x S$ into the amplitude ($\partial_{\mathrm{amp}}S$) and operator ($\partial_{\mathrm{op}} S$) derivatives,
\[
    \partial_x S = \partial_{\mathrm{amp}}S + \partial_{\mathrm{op}} S,
\]
and realizing that the former commutes with $S$, we can compactly write the left-hand side of Eq.~\ref{covar-cc-gov} as
\begin{equation}
    e^{-S} \left ( \partial_x \: e^{S} \right ) = \frac{\partial_{\mathrm{amp}} S}{\partial x}+ S_x, \quad
    S_x = \int_0^1 dy \: e^{-yS}  \frac{ \partial_{\mathrm{op}} S}{ \partial x}   e^{yS},
\end{equation}
where the integration over $y$ in the second equation can be carried out analytically, as explained in Eq.~\ref{wilcox-demo-2}. With these details, Eq.~\ref{covar-cc-gov} can be further simplified as
\begin{subequations}
    \label{covar-cc-gov2}
    \begin{align}
        \frac{\partial_{\mathrm{amp}} S}{\partial \alpha}\: \vert \Psi_0 \rangle &= \left [ \frac{1}{2} \left ( e^{-S} N e^{S} - N \right ) - S_{\alpha} \right ] \: \vert \Psi_0 \rangle
        \\
        \frac{\partial_{\mathrm{amp}} S}{\partial \beta}\:  \vert \Psi_0 \rangle &= -\left [ \frac{1}{2} \left ( e^{-S} H e^{S} - H_0 \right ) + S_{\beta} \right ] \: \vert \Psi_0 \rangle,
    \end{align}
\end{subequations}
which can then be left-projected with various thermal quasiparticle states to yield a set of differential equations governing the evolution of the $s$-amplitudes in the chemical potential - temperature or $\alpha$-$\beta$ space. Complete expressions for the CCSD evolution equations are included in Appendix \ref{app1:eqns}.

\begin{figure*}[!ht]
    \subfloat[2-site, $U/t=1$\label{fig:2SitesHalfFilling}]{
        \centering
        \includegraphics[width=0.48\textwidth]{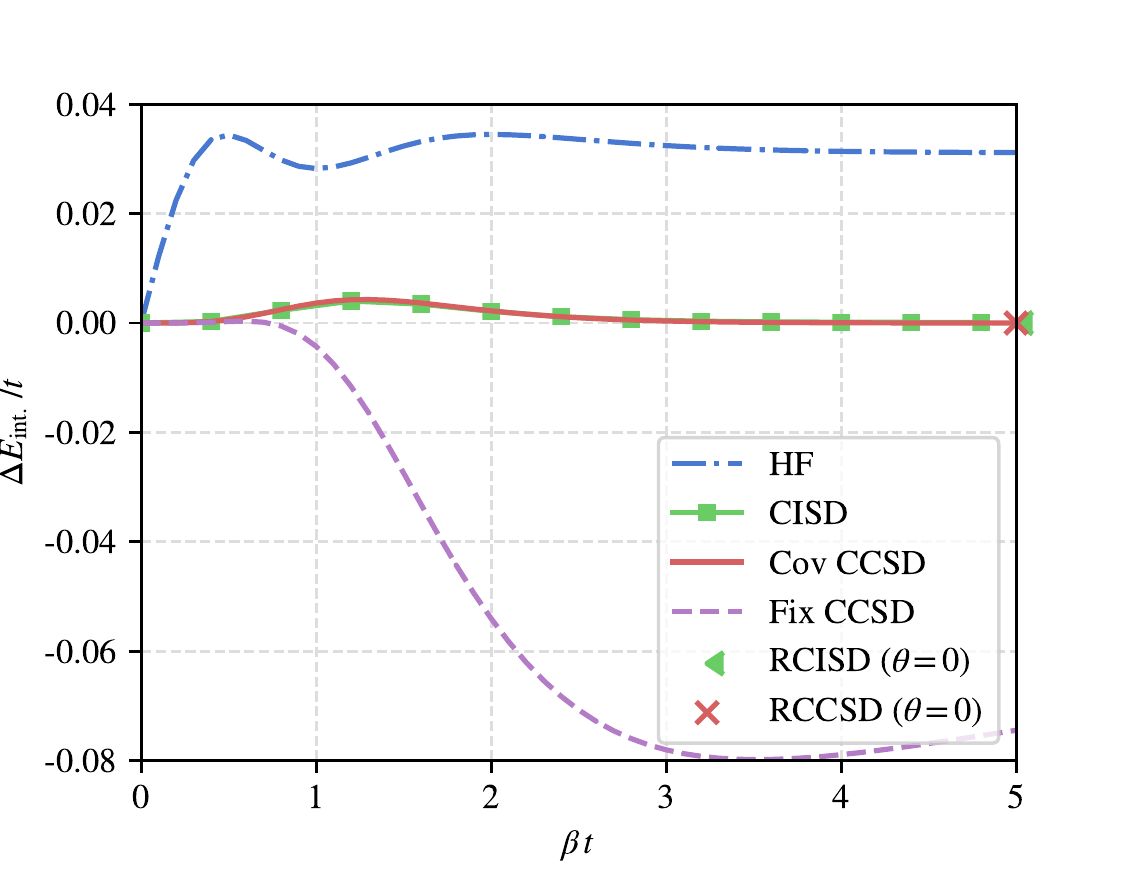}
    }
    \subfloat[6-site, $U/t=2$\label{fig:6SitesHalfFilling}]{
        \centering
        \includegraphics[width=0.48\textwidth]{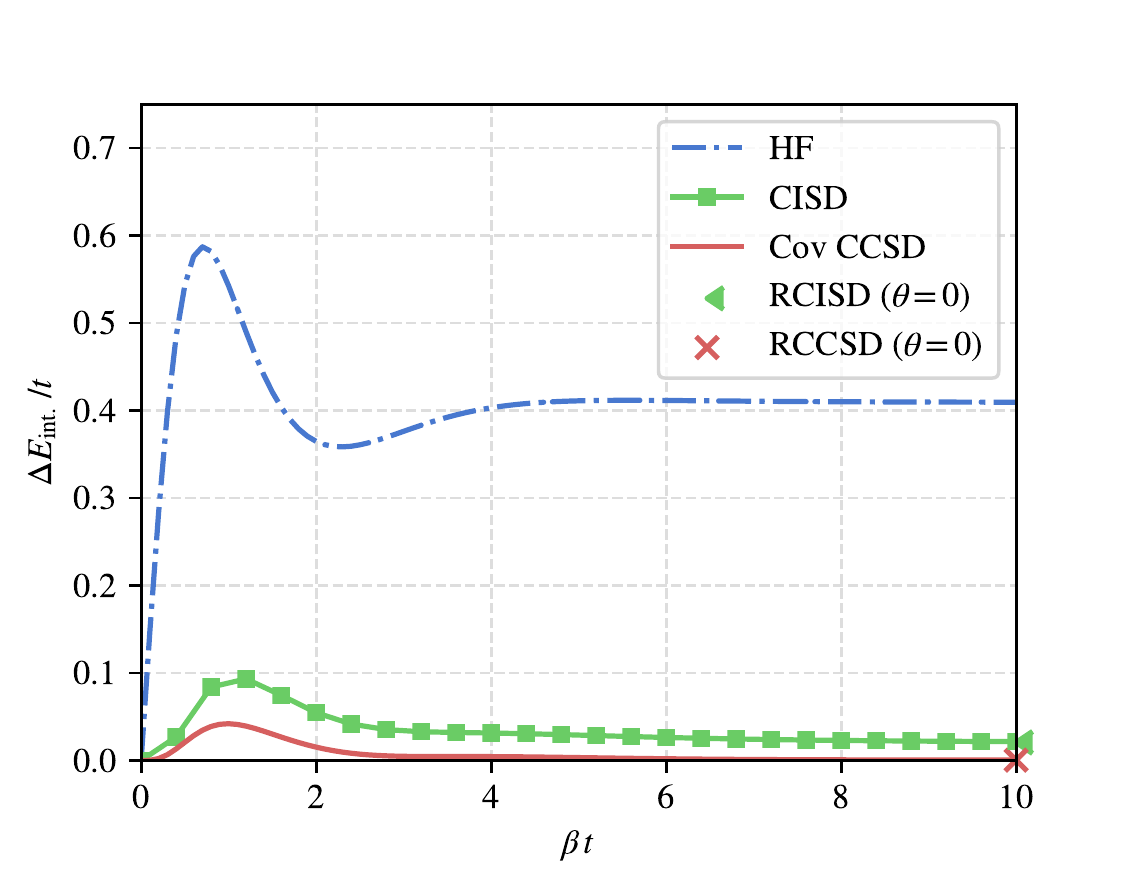}
    }
    \caption{Error in internal energy for thermal HF, covariant CISD, fixed-reference and covariant thermal CCSD for (a) two-site, and (b) six-site Hubbard models with $U/t = 1,2$ respectively at half filling on average.}
    \label{fig:HubbardHalfFilling}
\end{figure*}

\begin{figure*}[t]
    \subfloat[6-levels, $G=0.2$\label{fig:6LevelHalfFilling_G2}]{
        \centering
        \includegraphics[width=0.48\textwidth]{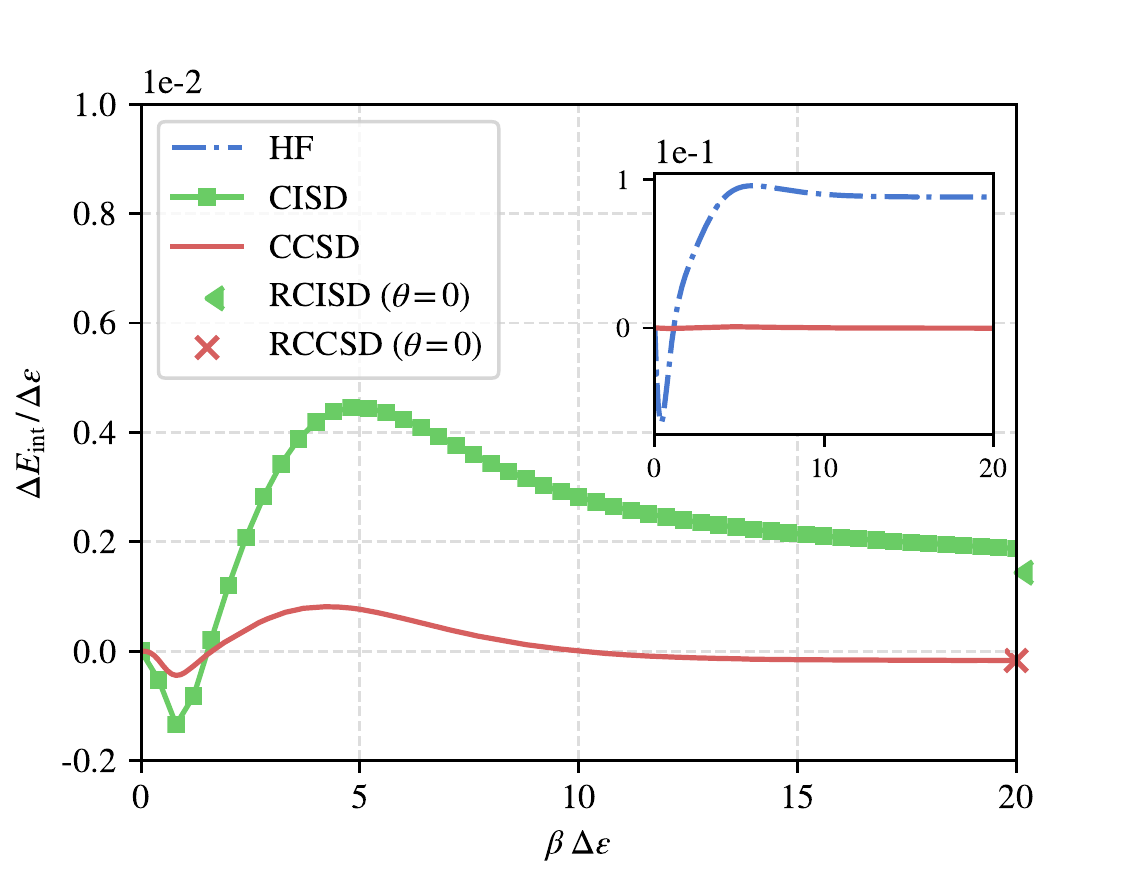}
    }
    \subfloat[6-levels, $G=0.5$\label{fig:6LevelHalfFilling_G5}]{
        \centering
        \includegraphics[width=0.48\textwidth]{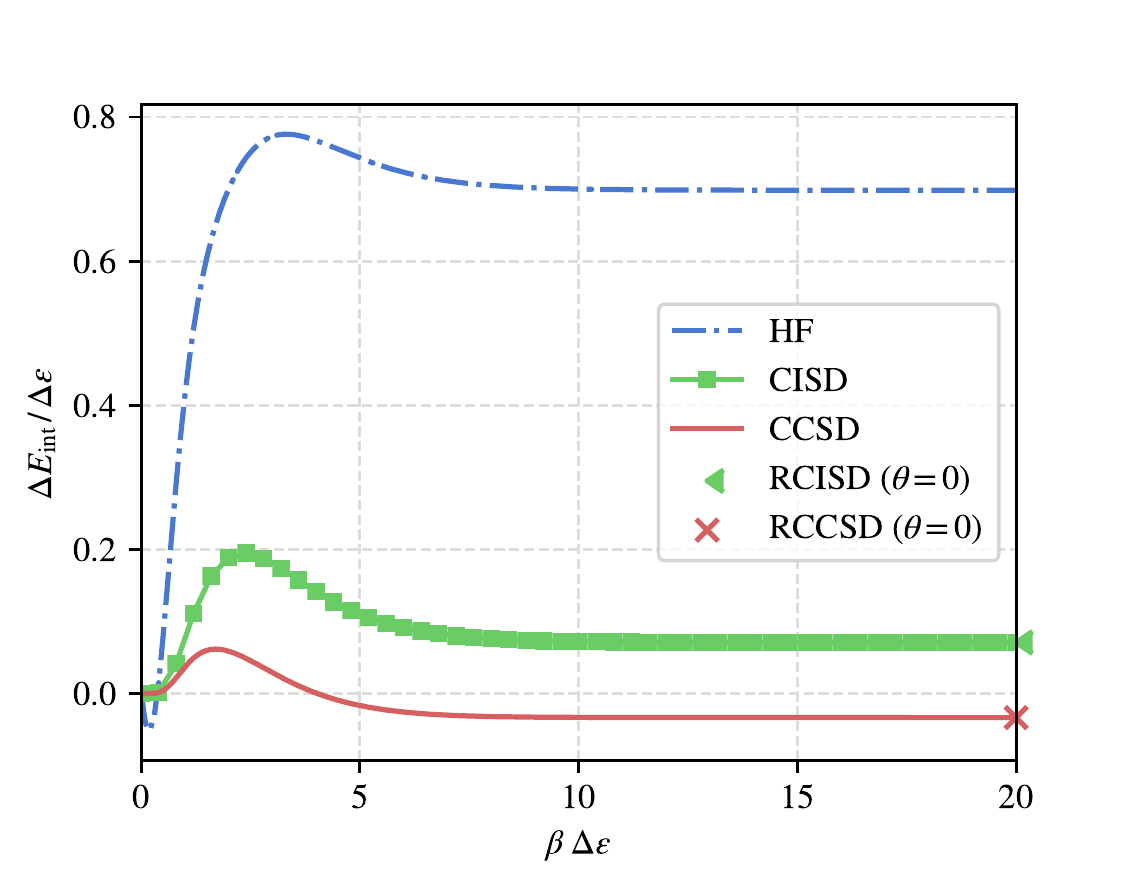}
    }
    \caption{Error in internal energy for thermal HF, covariant CISD and CCSD for the six-level pairing model with (a) $G = 0.2$, and (b) $G = 0.5$ respectively at half filling on average.}
    \label{fig:pairingHalfFilling}
\end{figure*}

\begin{figure}[t]
    \includegraphics[width=0.96\columnwidth]{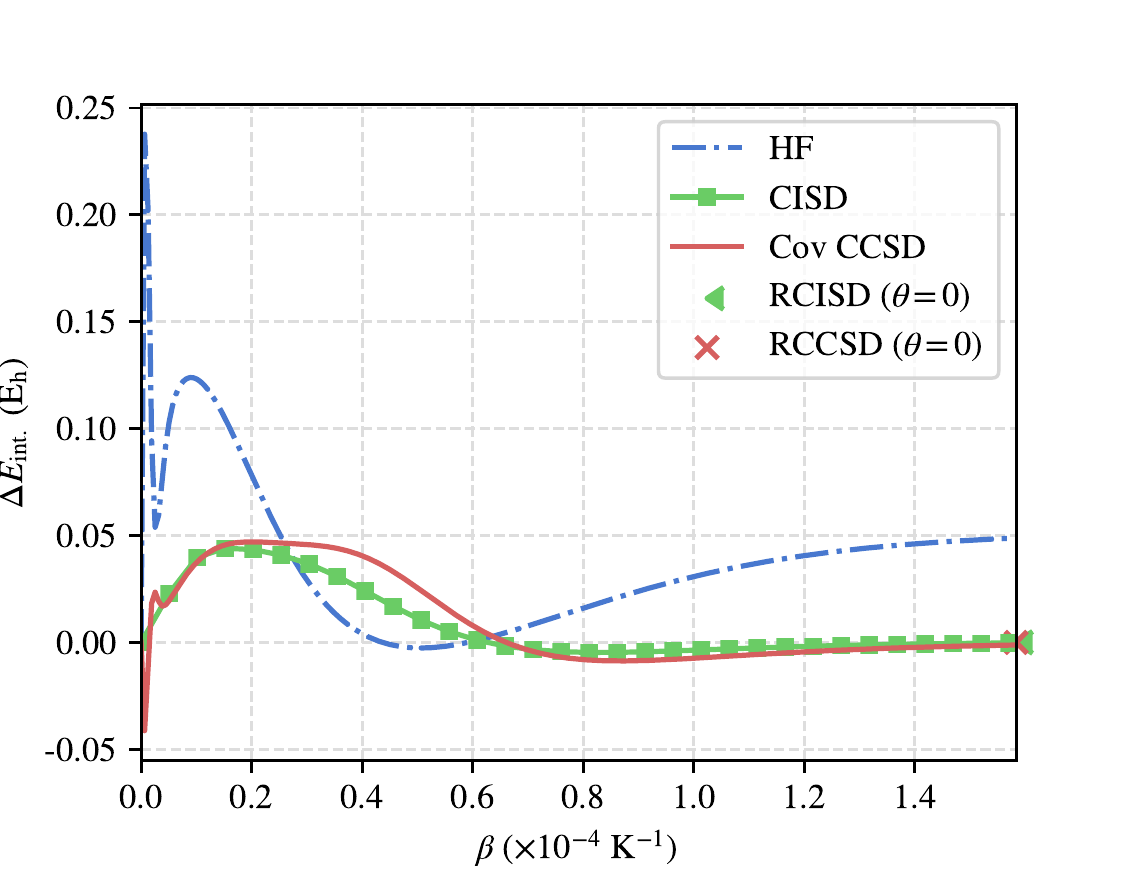}
    \caption{Error in internal energy for Be atom as function of temperature ($\theta$) in thermal HF, covariant CISD and CCSD as calculated in STO-3G basis.}
    \label{fig:Beryllium}
\end{figure}

\section{Implementation}
\subsection{Integration}
An explicit Runge-Kutta method of order (4)5 with step-size control\cite{dormand_family_1980} has been employed for integrating the resulting set of differential equations, where step-size is adaptively modified so as to keep the relative error in evolution within $10^{-8}$. Furthermore, for applications that require a fixed average filling fraction, a tolerance of $10^{-5}$ in the number of electrons is used to find the target chemical potential.

\subsection{Choice of $H_0$}
As mentioned earlier, for the covariant thermal CC, a mean-field Hamiltonian $H_0 = \sum_p \ep_p a_p^\dagger a_p$ is used to construct the mean-field thermal reference state $\vert \Psi_0 \rangle$. There are several alternatives to pick this mean-field Hamiltonian -- in thermal HF theory,\cite{mermin_stability_1963, sokoloff_consequences_1967} one uses an $H_0$ that optimizes the applicable free energy, and it therefore depends on $\alpha$ and $\beta$. Such a choice, while it may yield an excellent mean-field thermal reference state, will result in convoluted expressions for the mean-field evolution as well as the operator derivatives, and is beyond the scope of current work.

In our implementation, we use zero-temperature ground-state HF, which may or may not preserve the ground-state symmetries, to construct the energy eigenvalues in $H_0$. Using an $\alpha$- and $\beta$-independent $H_0$ is analogous to the imaginary-time interaction picture formalism and simplifies our implementation since, (\emph{i}) an $\alpha$- and $\beta$-independent $H_0$ leads to clean analytical forms for the mean-field evolution operator as well as the $\alpha$- and $\beta$-derivatives of the thermal quasiparticle operators, and (\emph{ii}) $H_0$ being diagonal yields a straightforward thermal Bogoliubov transformation in Eq~\ref{hfb-transformation}. We note, however, that this choice is different from the one used in Ref.~\onlinecite{harsha_thermofield_2019}, where we use simply the one-electron Hamiltonian to construct $H_0$.

\section{\label{sec5}Results}

Armed with the working equations, we now proceed to present results for the application of the thermal CC in the covariant formalism, truncated at singles and doubles (CCSD), to various many-electron systems, \emph{viz.} the one-dimensional Hubbard model,\cite{hubbard_electron_1963} the pairing or the reduced BCS model, as well as chemical systems (atomic Beryllium and molecular $\mathrm{H}_2$).
In order to make correspondence with the canonical ground state limit, we present results for all of these systems with a fixed number of particles on average - at each $\beta$ grid-point, we evolve the thermal states in $\alpha$ to fix the average number of particles before evolving again in $\beta$.
We compare our results with full configuration interaction (FCI) results.

We first apply the CCSD methods to the one-dimensional Hubbard model with periodic boundary conditions. Having already presented results for thermal CI truncated to singles and doubles (CISD) in Ref.~\onlinecite{harsha_thermofield_2019}, this model system seems to be the right place to start comparing thermal CC with thermal HF and CI. The Hamiltonian is given by
\begin{equation}
    H = -t \sum_{\langle p,q \rangle,  \sigma} \left(c_{p,\sigma}^\dagger \, c_{q,\sigma} + \textrm{h.c.}\right) + U \, \sum_{p} n_{p,\uparrow} \, n_{p,\downarrow},
\end{equation}
where $\langle,\rangle$ denotes that the sum is carried over sites connected in the lattice, $t$ denotes the strength of the kinetic energy term, $U$ denotes the strength of the on-site Coulomb repulsion, and $n_{p,\sigma} = c_{p,\sigma}^\dagger \, c_{p,\sigma}$ is the number operator for lattice site $p$ and spin $\sigma$.  The ratio $U/t$ characterizes the correlation strength.

\begin{figure}[t]
    \includegraphics[width=0.96\columnwidth]{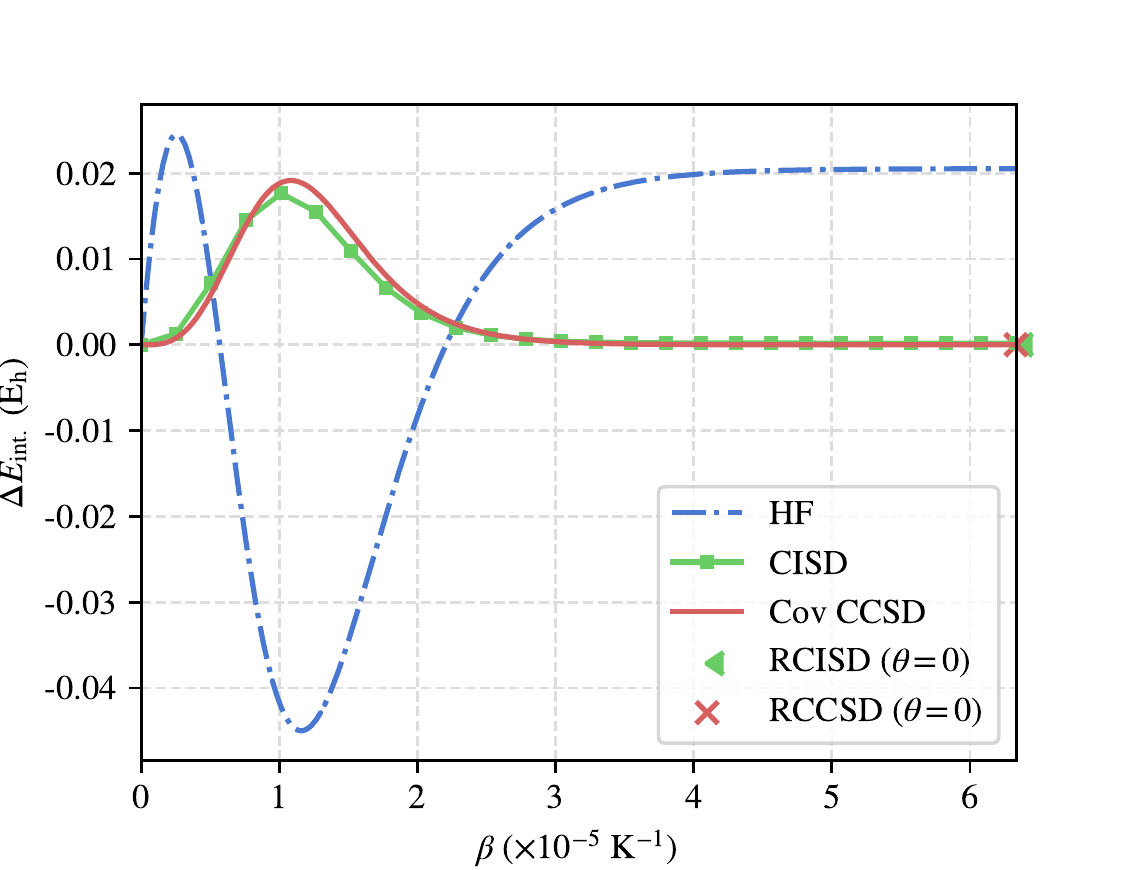}
    \caption{Error in internal energy for H$_2$ molecule at bond length of $0.74 \angstrom$ as function of temperature $\theta$ in thermal HF, covariant CISD and CCSD as calculated in STO-3G basis.}
    \label{fig:H2_sto3g}
\end{figure}

Figure~\ref{fig:2SitesHalfFilling} shows the temperature dependence of the error in internal energy for a two-site Hubbard model at half-filling on average with $U/t=1$ as computed by thermal HF, covariant CISD, thermal CCSD as well its fixed-reference formulation. Figure~\ref{fig:6SitesHalfFilling} presents the same for a six-site Hubbard model with $U/t=2$. For both cases, restricted HF (or RHF) eigenvalues and integrals have been used to construct $H_0$ and $H$ respectively. Thermal CCSD clearly outperforms CISD, especially for larger systems where CI is a less accurate wavefunction ansatz. Moreover, the covariant thermal CISD and CCSD go to the appropriate ground-state restricted CCSD in the limit $\beta \rightarrow \infty$ (or $\theta \rightarrow 0$, where $\theta = 1/\beta$ is the temperature). We note that while CCSD and CISD are exact in describing the ground state of the two-site Hubbard model, they are not exact at finite temperatures since we are working in the grand canonical ensemble.
The fixed-reference CCSD, on the other hand, performs poorly except for a small window around $\beta = 0$. Such a behaviour can be anticipated since this method uses the thermal reference corresponding to $\beta = 0$ as its starting point. Accordingly, for all other results that follow, we present only the covariant methods.

Next, we consider the reduced BCS or the pairing model, the Hamiltonian for which is given by
\begin{equation}
    H = \sum_p \epsilon_p N_p - G \sum_{p,q} P^\dagger_p P_q
\end{equation}
where $N_p$ counts the number of electrons, $\epsilon_p$ denotes the energy, and $P^\dagger_p$/$P_p$ respectively creates/annihilates a pair of electrons in the $p^{\mathrm{th}}$-level, while $G$ quantifies the attractive pair-hopping interaction. Here we choose the energy levels with a uniform spacing of 1 unit, i.e., $\Delta \epsilon = \epsilon_{p+1}-\epsilon_p = 1$.
Figure~\ref{fig:6LevelHalfFilling_G2} describes the temperature dependence of the error in internal energy for a six-level pairing model with $G = 0.2$ (weakly correlated) at average half-filling. Figure~\ref{fig:6LevelHalfFilling_G5} shows the same for $G = 0.5$ (near critical regime). Again, we use RHF eigenvalues and integrals to construct $H_0$ and $H$, and we see that the covariant thermal CCSD improves significantly over both the HF and CISD.
We also recover the zero-temperature ground-state limit for thermal CISD and CCSD.
Figures~\ref{fig:Beryllium} and \ref{fig:H2_sto3g} show similar trends for atomic Beryllium and molecular H$_2$ at bond length $0.74 \angstrom$ in STO-3G basis sets.

\begin{figure}[ht]
    \centering
    \includegraphics[width=0.96\columnwidth]{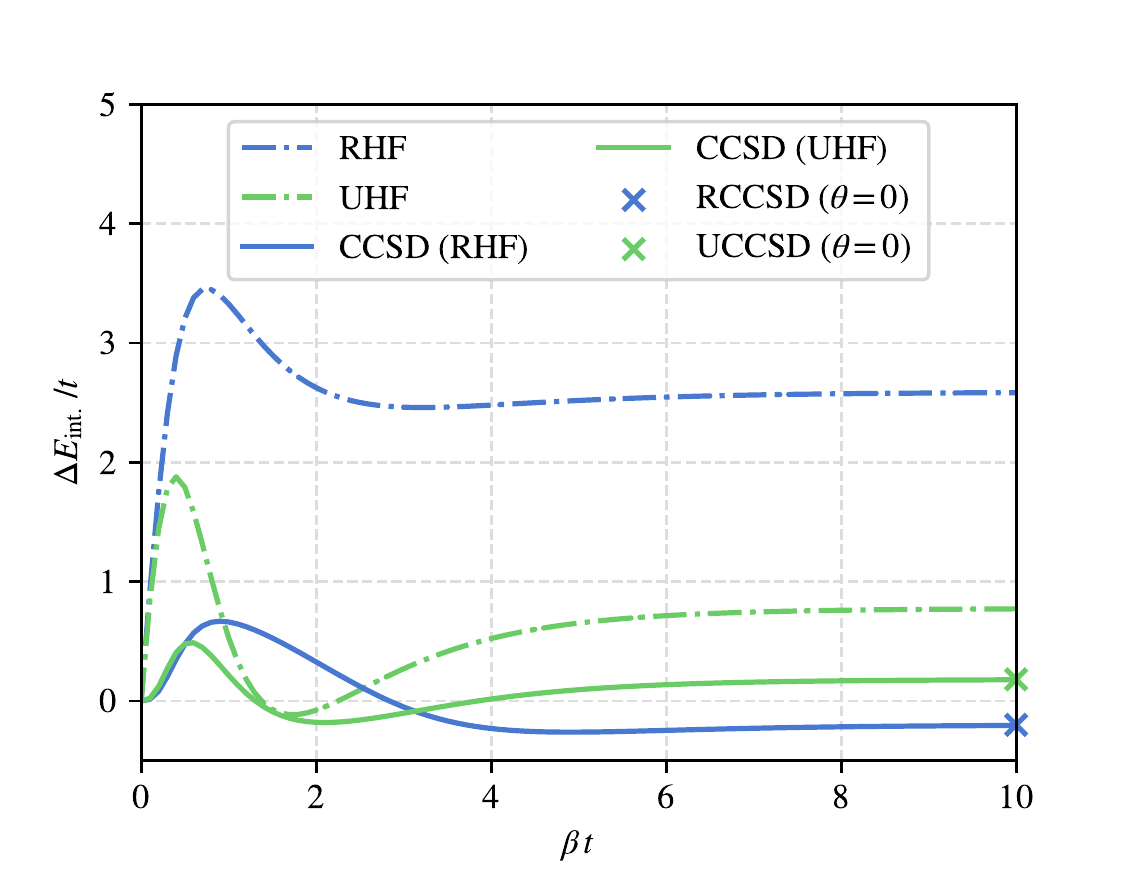}
    \caption{Error in internal energy for thermal HF and thermal CCSD, both based on RHF as well as UHF, for six-site Hubbard model with $U/t=5$ at half filling on average.}
    \label{fig:HubbardStrongCorr}
\end{figure}

\begin{figure}[tbh!]
    \centering
    \includegraphics[width=0.96\columnwidth]{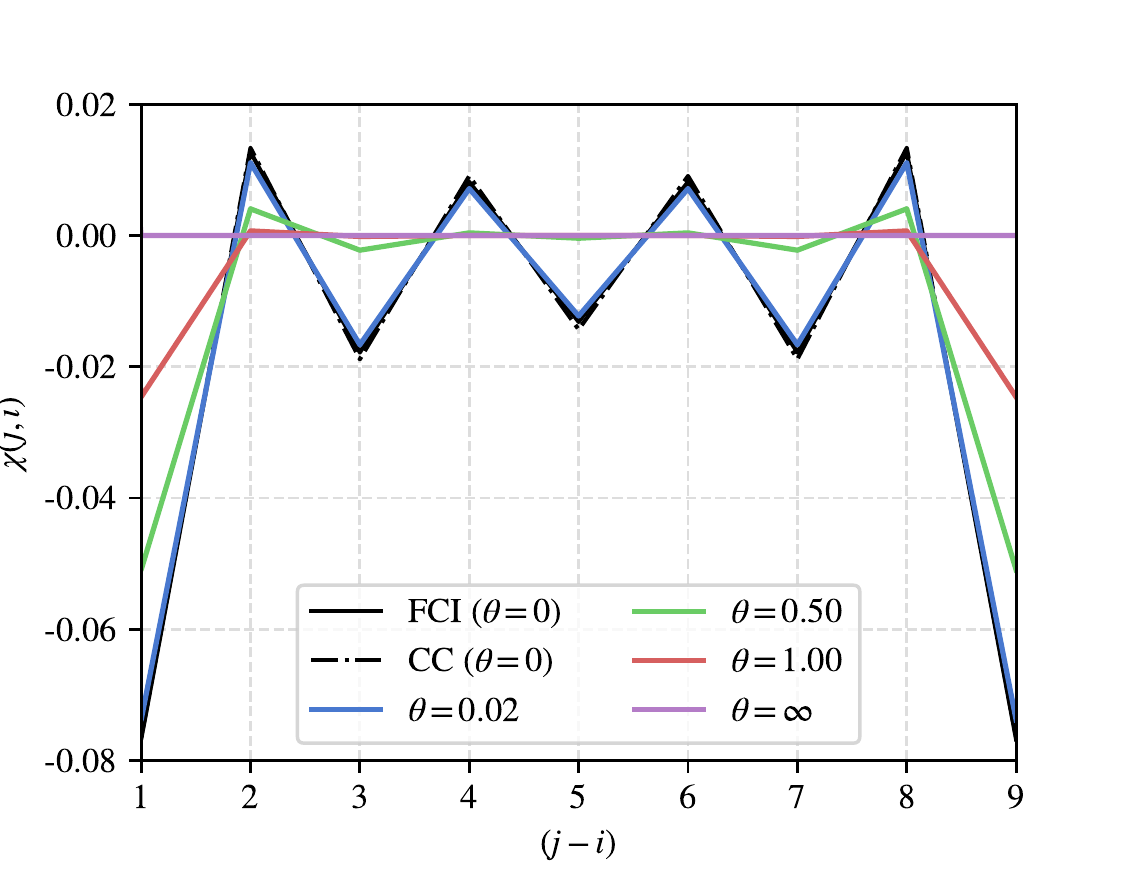}
    \caption{Trends in $z$-component spin-spin correlation function for the ten-site Hubbard model with $t=1,\, U=2$ at various temperatures $\theta$ and a fixed chemical potential, $\mu = 0.9$ (which ensures half-filling in the large $\beta$ limit), computed using thermal CCSD. Exact (FCI) and RCCSD results for the ground-state correlation functions are also included for reference.}
    \label{fig:spin-corr}
\end{figure}

In all of our applications so far, $H_0$ is constructed using RHF energy eigenvalues. Accordingly, in the zero-temperature limit, thermal HF approaches ground-state RHF and thermal CCSD approaches ground-state RCCSD. In strongly-correlated regime, where zero-temperature RCCSD fails to converge, the evolution of thermal CCSD also eventually diverges. Such an issue can be avoided by using unrestricted HF (or UHF) energy eigenvalues and integrals to construct $H_0$ and $H$ respectively. Figure~\ref{fig:HubbardStrongCorr} shows the performance of RHF and UHF based thermal CCSD internal energies against FCI results for six-site Hubbard model at half-filling on average with $U/t=5$.

In addition to the internal energy, we can also compute other physical properties and correlation functions at any temperature / chemical potential. In Figure~\ref{fig:spin-corr}, we show the $z$-component spin-spin correlation function
\begin{equation}
    \chi(i,j) = \langle S^z(i)\:S^z(j) \rangle
\end{equation}
for the ten-site Hubbard model with $U/t = 2$. Here, the expectation values are computed using linear-response density matrices and orbital relaxation effects have not been considered. As one would expect, at very high temperature $\theta$, there is no correlation between adjacent spins. As $\theta$ is reduced, the correlation appears and becomes maximal in the zero-temperature limit. Again, as with the internal energies, we see that the correlation function approaches the ground state CC in the limit $\beta \rightarrow \infty$.

\section{Conclusions}

We have demonstrated that the framework of thermofield dynamics can be exploited to formulate a finite-temperature coupled cluster theory.
We use the CCSD approximation to benchmark our method on various many-electron Hamiltonians and find that it performs substantially better than the thermal Hartree-Fock and thermal CI, just as one would expect for their ground-state counterparts.
This improved accuracy comes with the same asymptotic $\mathcal{O}(N^6)$ scaling ($N$ being the number of spin-orbitals or basis functions) as does standard quasiparticle CCSD, with a modestly larger pre-factor, though note that we must solve these equations at each grid point in the imaginary-time evolution.
We also observe that in the zero-temperature limit, thermal HF and CCSD approaches the appropriate ground-state HF and CCSD.

The thermofield based formulation of CC provides the basis for generalization of more sophisticated CC ansatze, e.g. unitary and multi-reference CC, to finite-temperatures, something that will be explored in future work.
Moreover, while CCSD with a perturbative treatment of triple excitations, or CCSD(T), has claimed the status of ``gold standard'' among ground-state methods for achieving chemical accuracy in weakly correlated systems, it is not clear how a similar notion can be defined for finite-temperature methods. More benchmark studies need to be performed to understand and improve accuracy of thermal CC.

\begin{acknowledgments}
        This work was supported by the U.S. Department of Energy, Office of Basic Energy Sciences, Computational and Theoretical Chemistry Program under Award No. DE-FG02-09ER16053. G.E.S. acknowledges support as a Welch Foundation Chair (No. C-0036).
\end{acknowledgments}

\appendix

\section{\label{app1:eqns}Thermal CCSD Equations}
When the cluster operator $S$ is truncated to singles and double excitation operators only, i.e.,
\[
    S = s_0 + \sum_{pq} s_{pq} a_p^\dagger \tilde{a}_q^\dagger + \frac{1}{4} \sum_{pqrs} s_{pqrs} a_p^\dagger a_q^\dagger \tilde{a}_s^\dagger \tilde{a}_r^\dagger
\]
the evolution equations (c.f. Eq.~\ref{covar-cc-gov}) for the CC amplitudes take the following form for the $\alpha$-evolution,
\begin{align}
    \frac{\partial s_0}{\partial \alpha} &= \frac{1}{2} \mathcal{R}_N^0 - S_\alpha^0
    \\
    \frac{\partial s_{pq}}{\partial \alpha} &= \frac{1}{2} \mathcal{R}_N^{pq} - S_\alpha^{pq},
    \\
    \frac{\partial s_{pqrs}}{\partial \alpha} &= \frac{1}{2} \mathcal{R}_N^{pqrs} - S_\alpha^{pqrs},
\end{align}
where $\mathcal{R}_N$ denotes the various CC residuals for the $\alpha$-evolution,
\begin{subequations}
    \begin{align}
        \mathcal{R}_N^0 &= \langle \Psi_0 \vert \left( e^{-S} N e^S  - N \right ) \vert \Psi_0 \rangle,
        \nonumber
        \\
        & = \sum_a x_a y_a s_{aa}
        \\
        \mathcal{R}_N^{pq} &= \langle \Psi_0 \vert \tilde{a}_q a_p \left( e^{-S} N e^S  -  N \right) \vert \Psi_0 \rangle,
        \nonumber
        \\
        & = ( x_p^2 - y_q^2 ) s_{pq}
        + \sum_a x_a y_a \left( s_{apaq} - s_{pa} s_{aq} \right)
        \\
        \mathcal{R}_N^{pqrs} &= \langle \Psi_0 \vert \tilde{a}_r \tilde{a}_s a_q a_p \left( e^{-S} N e^S  -  N \right) \vert \Psi_0 \rangle,
        \nonumber
        \\
        & = ( x_p^2 + x_q^2 - y_r^2 - y_s^2 ) s_{pqrs}
        \nonumber
        \\
        & \quad+ \frac{1}{2} \mathcal{P}(p q) \mathcal{P}(r s)
                \sum_a x_a y_a \left( s_{as} s_{pqar} - s_{pa} s_{aqrs} \right)
    \end{align}
\end{subequations}
and the operator-derivative terms are given by
\begin{subequations}
    \begin{align}
        S_\alpha^0 &= -\frac{1}{2} \sum_a x_a y_a s_{aa}
        \\
        S_\alpha^{pq} &= \frac{1}{2} \sum_a x_a y_a \left( s_{pa} s_{aq} + s_{paaq} \right)
        \\
        S_\alpha^{pqrs} &= - \frac{1}{4} \mathcal{P}(pq) \mathcal{P}(rs) \sum_a x_a y_a \left( s_{as} s_{pqar} - s_{pa} s_{aqrs} \right),
    \end{align}
\end{subequations}
where $x$ and $y$ are the thermal Bogoliubov parameters and the dummy indices $a$, $b$, $\ldots$ are summed over all the spin-orbitals.
Equations for the $\beta$-evolution can be obtained in a similar way,
\begin{align}
    \frac{\partial s_0}{\partial \beta} &= \frac{1}{2} \mathcal{R}_H^0 - S_\beta^0
    \\
    \frac{\partial s_{pq}}{\partial \beta} &= \frac{1}{2} \mathcal{R}_H^{pq} - S_\beta^{pq},
    \\
    \frac{\partial s_{pqrs}}{\partial \beta} &= \frac{1}{2} \mathcal{R}_H^{pqrs} - S_\beta^{pqrs},
\end{align}
where $\mathcal{R}_H$ denotes the various CC residuals for the $\beta$-evolution. Using the thermal Bogoliubov transformation described in Eq.~\ref{hfb-transformation}, a general two-body Hamiltonian,
\[
    H = \sum_p h_{pq} c_p^\dagger c_p + \frac{1}{4} \sum_{pqrs} u_{pqrs} c_p^\dagger c_q^\dagger c_s c_r,
\]
can be expressed in terms of thermal creation/annihilation operators, and takes the form
\begin{widetext}

    \begin{align}
        H &= h_0 + \sum_{ab} \left [
            h^{(11)}_{ab} \left( a^\dagger_a \tilde{a}^\dagger_b + \mathrm{h.c.} \right)
            + h^{(20)}_{ab} a_a^\dagger a_b + h^{(02)}_{ab} \tilde{a}_a^\dagger \tilde{a}_b
        \right]
        + \sum_{abcd} \left[
            h^{(221)}_{abcd} \left( a_a^\dagger a_b^\dagger \tilde{a}_d^\dagger \tilde{a}_c^\dagger + \mathrm{h.c.}\right)
            + h^{(222)}_{abcd} a_a^\dagger \tilde{a}_b^\dagger \tilde{a}_d a_c
        \right.
        \nonumber
        \\
        & \quad \quad \quad \quad \left.
            + h^{(31)}_{abcd} \left( a_a^\dagger a_b^\dagger \tilde{a}_c^\dagger a_d + \mathrm{h.c.} \right)
            + h^{(13)}_{abcd} \left( a_a^\dagger \tilde{a}_b^\dagger \tilde{a}_c^\dagger \tilde{a}_d + \mathrm{h.c.} \right)
            + h^{(40)}_{abcd} a_a^\dagger a_b^\dagger a_d a_c + h^{(04)}_{abcd} \tilde{a}_a^\dagger \tilde{a}_b^\dagger \tilde{a}_d \tilde{a}_c
        \right]
    \end{align}
    where we use $h_0$, $h^{(11)}$, etc. to denote the effective matrix elements of the general quasiparticle Hamiltonian ($h^{(11)}$ is associated with operators that contain a non-\textit{tilde} and a \textit{tilde} quasiparticle each, $h^{(20)}$ with two non-\textit{tilde} quasiparticle operators, and so on), which are given by
    \begin{equation}
        h_0 = \sum_a y_a^2 h_{aa} + \frac{1}{2} \sum_{ab} y_a^2 y_b^2 u_{abab}
    \end{equation}
    \begin{equation}
        h^{(11)}_{ab} = x_a y_b f_{ab}, \quad h^{(20)}_{ab} = x_a x_b f_{ab}, \quad h^{(02)}_{ab} = -y_a y_b f_{ab},
        \quad \mathrm{with} \quad
        f_{ab} = \delta_{ab} h_{ab} + \sum_c y_c^2 u_{acbc}
    \end{equation}
    \begin{align}
        h^{(221)}_{abcd} = \frac{1}{4} x_a x_b y_c y_d u_{abcd},
        \quad
        h^{(222)}_{abcd} &= x_a x_c y_b y_d u_{adbc},
        \quad
        h^{(31)}_{abcd} = - \frac{1}{2} x_a x_b y_c x_d u_{abcd},
        \nonumber
        \\
        h^{(13)}_{abcd} = -\frac{1}{2} x_a y_b y_c y_d u_{adbc},
        \quad
        h^{(40)}_{abcd} &= \frac{1}{4} x_a x_b x_c x_d u_{abcd},
        \quad
        h^{(04)}_{abcd} = \frac{1}{4} y_a y_b y_c y_d u_{abcd}.
    \end{align}
    In obtaining the above expressions, we have assumed real matrix elements in the Hamiltonian.
    The residuals can then be expressed compactly in terms of the effective Hamiltonian matrix elements,
    \begin{subequations}
        \begin{align}
            \mathcal{R}_H^0 &= \langle \Psi_0 \vert \left( e^{-S} H e^S  - H_0 \right ) \vert \Psi_0 \rangle,
            \nonumber
            \\
            & = h_0 - \sum_a y_a^2 \ep_a + \sum_{ab} h^{(11)}_{ab} s_{ab} + \sum_{abcd} \left( 2 s_{ac} s_{bd} + s_{abcd} \right) h^{(221)}_{abcd}
            \\
            \mathcal{R}_H^{pq} &= \langle \Psi_0 \vert \tilde{a}_q a_p \left( e^{-S} H e^S  -  H_0 \right) \vert \Psi_0 \rangle,
            \nonumber
            \\
            &= h^{(11)}_{pq} - \delta_{pq} \ep_p x_p y_p + \sum_a \left( h^{(02)}_{aq} s_{pa} + h^{(20)}_{ap} s_{aq} \right)
            - \sum_{ab} \left( h^{(11)}_{ab} \left ( s_{aq} s_{pb} + s_{pabq} \right ) - h^{(222)}_{pqab} s_{ab} \right )
            \nonumber
            \\
            & \quad + \sum_{abc} \left( h^{(13)}_{abcq} \left( 2 s_{ab} s_{pc} + s_{apbc} \right) - h^{(31)}_{abcp} \left( 2 s_{ac} s_{bq} + s_{abcq} \right) \right)
            - 2 \sum_{abcd} h^{(221)}_{abcd} \left (
                2 s_{ac} ( s_{bq} s_{pd} + s_{pbdq} ) - s_{aq} s_{bpcd} - s_{pc} s_{abdq}
            \right )
            \\
            \mathcal{R}_H^{pqrs} &= \langle \Psi_0 \vert \tilde{a}_r \tilde{a}_s a_q a_p \left( e^{-S} H e^S  -  H_0 \right) \vert \Psi_0 \rangle,
            \nonumber
            \\
            &= \mathcal{P}(pq) \mathcal{P}(rs) \left [
                h^{(221)}_{pqrs} + \sum_a \left(
                    \frac{1}{2}\left( h^{(02)}_{ar} s_{pqas} + h^{(20)}_{ap} s_{aqrs}\right) + h^{(13)}_{prsa} s_{qa} + h^{(31)}_{pqsa}s_{ar}
                \right)
            \right.
            \nonumber
            \\
            & \left.
                + \frac{1}{2} \sum_{ab} \left(
                    h^{(04)}_{abrs} ( 2 s_{pa} s_{qb} + s_{pqab} ) + h^{(40)}_{abpq} ( 2 s_{ar} s_{bs} + s_{abrs} )
                    - h^{(11)}_{ab} ( s_{ar} s_{pqbs} + s_{pb} s_{aqrs} ) - 2 h^{(222)}_{prab} ( s_{qabs} + s_{as} s_{qb} )
                \right)
            \right.
            \nonumber
            \\
            & \left.
                + \sum_{abc} \left(
                    h^{(13)}_{abcr} \left(
                        s_{ab} s_{pqcs} + \frac{1}{2} s_{as}( 2 s_{pb} s_{qc} + s_{pqbc} )  - 2 s_{pb} s_{aqcs}
                    \right)
                    + h^{(31)}_{abcq} \left(
                        s_{ac} s_{bprs} + \frac{1}{2} s_{pc} ( 2 s_{ar} s_{bs} + s_{abrs} ) - 2 s_{ar} s_{bpcs}
                    \right)
                \right)
            \right.
            \nonumber
            \\
            & \left.
                - \sum_{abcd} h^{(221)}_{abcd} \left(
                    2 s_{ac} ( s_{br} s_{pqds} + s_{pd} s_{bqrs} )
                    - \frac{1}{2} s_{ar} s_{bs} ( 2 s_{pc} s_{qd} + s_{pqcd} ) + 4 s_{ar} s_{pc} s_{bqds}
                \right.
            \right.
            \nonumber
            \\
            & \left.
                \left.
                    - \frac{1}{4} s_{abrs} ( 2 s_{pc} s_{qd} + s_{pqcd} ) + ( s_{abcr} s_{pqds} + s_{apcd} s_{bqrs} )
                    + 2 s_{bqdr} s_{apcs}
                \right)
            \right],
        \end{align}
    \end{subequations}
\end{widetext}
where we have used the Baker-Campbell-Hausdorff expansion to simplify the similarity transformation, i.e.
\begin{equation}
  e^{-S} H e^{S} = H + [H,S] + \frac{1}{2!} [[H,S],S] + \ldots.
\end{equation}
For a two-body Hamiltonian, with $S$ truncated to at most double quasiparticle excitations, this expansion truncates at fourth order. Diagrammatic expressions for these equations can also be formulated along similar lines as Bogoliubov coupled cluster methods.\cite{signoracci_ab_2015}
The operator-derivative terms in the $\beta$-evolution are given by
\begin{subequations}
    \begin{align}
        S_\beta^0 &= \frac{1}{2} \sum_a \ep_a x_a y_a s_{aa}
        \\
        S_\beta^{pq} &= -\frac{1}{2} \sum_a \ep_a x_a y_a \left( s_{pa} s_{aq} + s_{paaq} \right)
        \\
        S_\beta^{pqrs} &= \frac{1}{4} \mathcal{P}(pq) \mathcal{P}(rs) \sum_a \ep_a x_a y_a \left( s_{as} s_{pqar} - s_{qa} s_{aprs} \right) .
    \end{align}
\end{subequations}

\section{\label{app2:fixed}Fixed-reference formulation}
In the fixed-reference formalism, which is quite analogous to the Schr\"odinger picture approach, we choose the state $\vert \mathbb{I} \rangle$ as the zeroth order approximation to the thermal state. However, as we have discussed in Ref.~\citenum{harsha_thermofield_2019}, one can chose any value of chemical potential $\alpha_0$ and temperature $\beta_0$ to construct the thermal reference state. It is merely a matter of comfort to use $\alpha_0 = 0 = \beta_0$ as the corresponding initial conditions for the cluster amplitudes are trivial. For $\vert \mathbb{I} \rangle$ as our choice of reference, it is convenient to redefine the thermal state as
\begin{equation}
    \vert \psi (\alpha, \beta) \rangle = e^{\alpha N-\beta H} \vert \mathbb{I} \rangle,
\end{equation}
so that the thermal expectation value of any physical quantity $A$ becomes
\begin{subequations}
    \begin{align}
        \langle A \rangle &= \frac{ \langle \mathbb{I} \vert \: A \: e^{-\beta H} \: \vert \mathbb{I} \rangle }{ \langle \mathbb{I} \vert \: e^{-\beta H} \: \vert \mathbb{I} \rangle }, \\
        &= \frac{ \langle \mathbb{I} \vert \: A \: \vert \psi (\alpha, \beta) \rangle }{ \langle \mathbb{I} \vert \psi (\alpha, \beta) \rangle },
    \end{align}
\end{subequations}
and a better \textit{bra} is no longer required. Correspondingly, the governing imaginary time Schr\"odinger equations become
\begin{subequations}
    \label{beta-mu-schro-fixref}
    \begin{align}
        \frac{ \partial }{\partial \beta} \vert \psi (\alpha, \beta) \rangle &= -H \vert \psi (\alpha, \beta) \rangle.
        \label{beta-schro-eq-fixref}
        \\
        \frac{ \partial }{\partial \alpha} \vert \psi (\alpha, \beta) \rangle &= N \vert \psi (\alpha, \beta) \rangle.
        \label{alpha-schro-eq-fixref}
    \end{align}
\end{subequations}
The thermal state for a given chemical potential $\alpha$ at inverse temperature $\beta$ can then be written as an exponential coupled cluster wavefunction
\begin{equation}
    \vert \Psi (\alpha, \beta) \rangle = e^{T (\alpha,\beta)} \: \vert \mathbb{I} \rangle,
\end{equation}
where the cluster operator $T (\alpha,\beta)$ builds correlation atop $\vert \mathbb{I}\rangle $. With this CC wavefunction ansatz, the finite-temperature expectation value of any physical quantity $A$ becomes
\begin{equation}
    \langle A \rangle = \langle \mathbb{I} \vert A e^{T} \vert \mathbb{I} \rangle e^{-t_0} = \langle \mathbb{I} \vert e^{-T} A e^{T} \vert \mathbb{I} \rangle.
    \label{fixref-expectation}
\end{equation}
The state $\vert \mathbb{I} \rangle$ is annihilated by thermal quasiparticle operators $a_p$ and $\tilde{a}_p$ corresponding to $x_p = y_p = 1/\sqrt{2}$ in the Bogoliubov transformation in Eq.~\ref{hfb-transformation}. We will refer these field operators as
\[
    d_p,\: d^\dagger_p,\: \tilde{d}_p,\: \tilde{d}^\dagger_p.
\]
Therefore, the cluster operator $T$ can be expressed as
\begin{equation}
    T = t_0
    + \sum_{p,q} t_{pq} d^\dagger_p \tilde{d}^\dagger_p
    + \frac{1}{(2!)^2} \sum_{p,q,r,s} t_{pqrs} d^\dagger_p d^\dagger_q \tilde{d}^\dagger_s \tilde{d}^\dagger_r
    + \ldots,
    \label{fixref-cluster}
\end{equation}
where the $\alpha$- and the $\beta$-dependence is carried by the cluster amplitudes.
These cluster amplitudes are found by integrating the imaginary time Schr\"odinger Eq.~\ref{beta-mu-schro-fixref}, which, upon substituting the wavefunction ansatz of Eq.~\ref{fixref-cluster}, gives the following working equation
\begin{subequations}
    \label{fixref-cc-gov}
    \begin{align}
        \frac{\partial T}{\partial \alpha} \: \vert \mathbb{I} \rangle &=
        e^{-T} N e^{T} \: \vert \mathbb{I} \rangle,
        \\
        \frac{\partial T}{\partial \beta} \: \vert \mathbb{I} \rangle &=
        - e^{-T} H e^{T} \: \vert \mathbb{I} \rangle.
    \end{align}
\end{subequations}
Like conventional ground-state CC, Eq.~\ref{fixref-cc-gov} can be left-projected with the ground and excited slater determinants to yield the evolution equations for the amplitudes,
\begin{subequations}
    \label{t-amp-eq}
    \begin{align}
        \frac{\partial t_0}{\partial \beta} &= - \frac{1}{\mathcal{Z}_{\mathbb{I}}} \langle \mathbb{I} \vert \: e^{-T} H e^{T} \: \vert \mathbb{I} \rangle,\\
        \frac{\partial t_{pq}}{\partial \beta} &= - \frac{1}{\mathcal{Z}_{\mathbb{I}}} \langle \mathbb{I} \vert \: \tilde{d}_q d_p \vert \: e^{-T} H e^{T} \: \vert \mathbb{I} \rangle,\\
        \frac{\partial t_{pqrs}}{\partial \beta} &= - \frac{1}{\mathcal{Z}_{\mathbb{I}}} \langle \mathbb{I} \vert  \: \tilde{d}_r \tilde{d}_s d_q d_p \: e^{-T} H e^{T} \: \vert \mathbb{I} \rangle,
    \end{align}
\end{subequations}
and so on, where $\mathcal{Z}_{\mathbb{I}} = \langle \mathbb{I} \vert \mathbb{I} \rangle$. Similar equations can be derived for evolution along $\alpha$. These equations can be integrated starting from $\beta = 0$ (or any other value), where the initial values of the amplitudes are known, to the required inverse temperature and chemical potential. Here, since $\vert \mathbb{I} \rangle $ is exact at $\alpha, \beta = 0$, we have the initial conditions,
\[
    t_0 = 0, \: t_{pq} = 0, \: t_{pqrs} = 0, \: \ldots.
\]

\bibliography{ThermalCC}

\end{document}